\documentclass{iau}
\usepackage{graphicx}

\title[Gaia: The Astrometry Revolution] 
{Gaia: The Astrometry Revolution}

\author[A. Sozzetti et al.]   
{A. Sozzetti$^1$, 
M. Bonavita$^2$, S. Desidera$^3$, R. Gratton$^3$, \and M.G. Lattanzi$^1$}

\affiliation{$^1$INAF - Osservatorio Astrofisico di Torino - Via Osservatorio 20, I-10025 Pino Torinese (Italy)\\email: {\tt sozzetti@oato.inaf.it}\\
$^2$The University of Edinburgh, Royal Observatory, Blackford Hill, Edinburgh EH9 3HJ, UK\\
$^3$INAF - Osservatorio Astronomico di Padova - Vicolo dell'Osservatorio 5, I-35122 Padova (Italy)
}

\pubyear{2015}
\volume{314}  
\pagerange{???--???}
\jname{Young Stars \& Planets Near the Sun}
\editors{J. H. Kastner, B. Stelzer, \& S. A. Metchev, eds.}
\begin{document}

\maketitle

\begin{abstract}
The power of micro-arcsecond ($\mu$as) astrometry is about to be unleashed. ESA's Gaia mission, 
now headed towards the end of the first year of routine science operations, will soon fulfil its promise 
for revolutionary science in countless aspects of Galactic astronomy and astrophysics. The potential of Gaia 
position measurements for important contributions to the astrophysics of planetary systems is huge. We focus 
here on the expectations for detection and improved characterization of 'young' planetary systems in the neighborhood 
of the Sun using a combination of Gaia $\mu$as astrometry and direct imaging techniques. 

\end{abstract}

\firstsection 
\section{Introduction}

The study of nearby young moving groups (NYMGs) in the vicinity ($d\lesssim100-200$ pc) of the Sun is of particular 
relevance to improve our understanding of many a key issue related to the early evolutionary stages of stars, circumstellar 
disks, and planetary systems. NYMGs are excellent laboratories indeed, as their precise and accurate identification, the 
determination of clean samples of bona-fide members (down to the sub-stellar regime), their origin, age, 
distance from Earth, and multiplicity properties (see, e.g., the review and contributed papers by Kastner, Mamajek, Pinsonneault, 
Elliott and Faherty, this volume) are keys to a) provide new insights into the early evolution of low- to intermediate-mass stars (see, e.g., 
the review and contributed papers by Feiden, Baraffe, Matt and Kraus, this volume), 
b) shed light on protoplanetary disk sculpting and dispersal (see, e.g., the review papers by Cieza, Birnstiel and Kospal, this volume), 
and c) globally understand giant and terrestrial planet formation, 
and the transition region between giant planets and brown dwarfs (see, e.g., the review and contributed papers by Mordasini, Chauvin, 
Marley, Allers and Youdin, this volume).

While data collected with a variety of techniques over the past two decades has allowed to make significant progress in the field 
thanks to the identification of hundreds of young nearby stars, the impact of new and future facilities on the study of NYMGs
and their members is expected to be revolutionary. Quantum leaps in our knowledge (surprises included!) will come from the next 
generation of large-scale synoptic surveys in the visible (e.g., http://www.lsst.org/), sub-millimeter and milliter observations 
(e.g., Wilner, this volume), wide-field spectroscopic surveys at visible wavelengths (e.g., Martell, this volume), and 
direct imaging surveys in the visible and near-infrared (e.g., Marois, this volume). In this respect, the promise for huge progress 
in all the above mentioned key areas will soon be achieved when data will become available for ESA's billion star surveyor: Gaia. 


\section{Gaia: The Dawn of the Age of $\mu$as Astrometry}

Gaia (http://www.cosmos.esa.int/gaia) is the first experiment set to
demonstrate single-epoch measurement accuracies $\sigma_\mathrm{A}\approx20$ $\mu$as for bright stars (see, e.g. the review on the mission, 
including its payload, by de Bruijne et al. 2010). The mission is now entering its second year of science operations at L2, 
after its successful launch in December 2013. Gaia's exquisite astrometric sensitivity will allow to unravel the formation history, 
evolution, structure, and dynamics of the Milky Way, through measurements of the positions, motions, distances, and astrophysical parameters 
of the brightest 1,000 million stars in the sky (Perryman et al. 2001). At the end of an over six-months long commissioning phase, a number of issues arose, 
which pose a challenge to data reduction pipelines in order to demonstrate pre-launch performance estimates. First, significant stray light levels 
were identified. Second, the transmission of the optics slowly degrades with time, as a result of contamination by water ice. Finally, 
the intrinsic instability of the basic angle which separates the lines of sight of the two telescopes is larger than expected. While a number 
of additional calibration procedures are being put in place in order to cope with, mitigate, and possibly remove these undesired effects, 
the impact on the astrometric performance of Gaia is still under evaluation, but the hope is that such complications will still fit within 
the 20\% margin included in pre-launch calculations (for a recent review of the issue, see de Bruijine et al. 2015). 

\section{Astrometric Planet Detection within the Gaia Pipeline}

High-precision global astrometry with Gaia is poised to enable the detection of planetary-mass companions around stars in the Solar neighborhood. 
The determination of the astrometric orbits of extrasolar planets will be obtained as part of the non-single star (NSS) treatment within Coordination 
Unit 4 (Object Processing) of the Gaia Data Processing and Analysis Consortium. CU4 will tackle the NSS problem by attempting to derive, based on the available spectroscopic
and photometric Gaia data, spectroscopic and/or eclipsing binary solutions. For astrometry, a cascade of increasingly more complex models will describe 
the data in terms of solutions containing derivatives of the stellar proper motion, accounting for variability induced
motion, all the way to fully Keplerian astrometric orbital solutions, including where appropriate multiple companions. 
A Development Unit (DU437) has been specifically devoted to the modelling of the astrometric signals produced by planetary systems. 
The DU is composed of several tasks, which implement multiple robust procedures for (single and multiple) astrometric orbit fitting (such as Markov Chain Monte Carlo 
and genetic algorithms) and the determination of the degree of dynamical stability of multiple-component systems. This robust approach to orbit modeling is expected 
to allow coping with the complexities inherent to adjusting large, non-linear models to the data (e.g., Sozzetti 2005, 2014)

\section{Planets Around Young Stars: The Gaia Potential}

\begin{figure}
\centering
$\begin{array}{cc}
\includegraphics[width=0.50\textwidth]{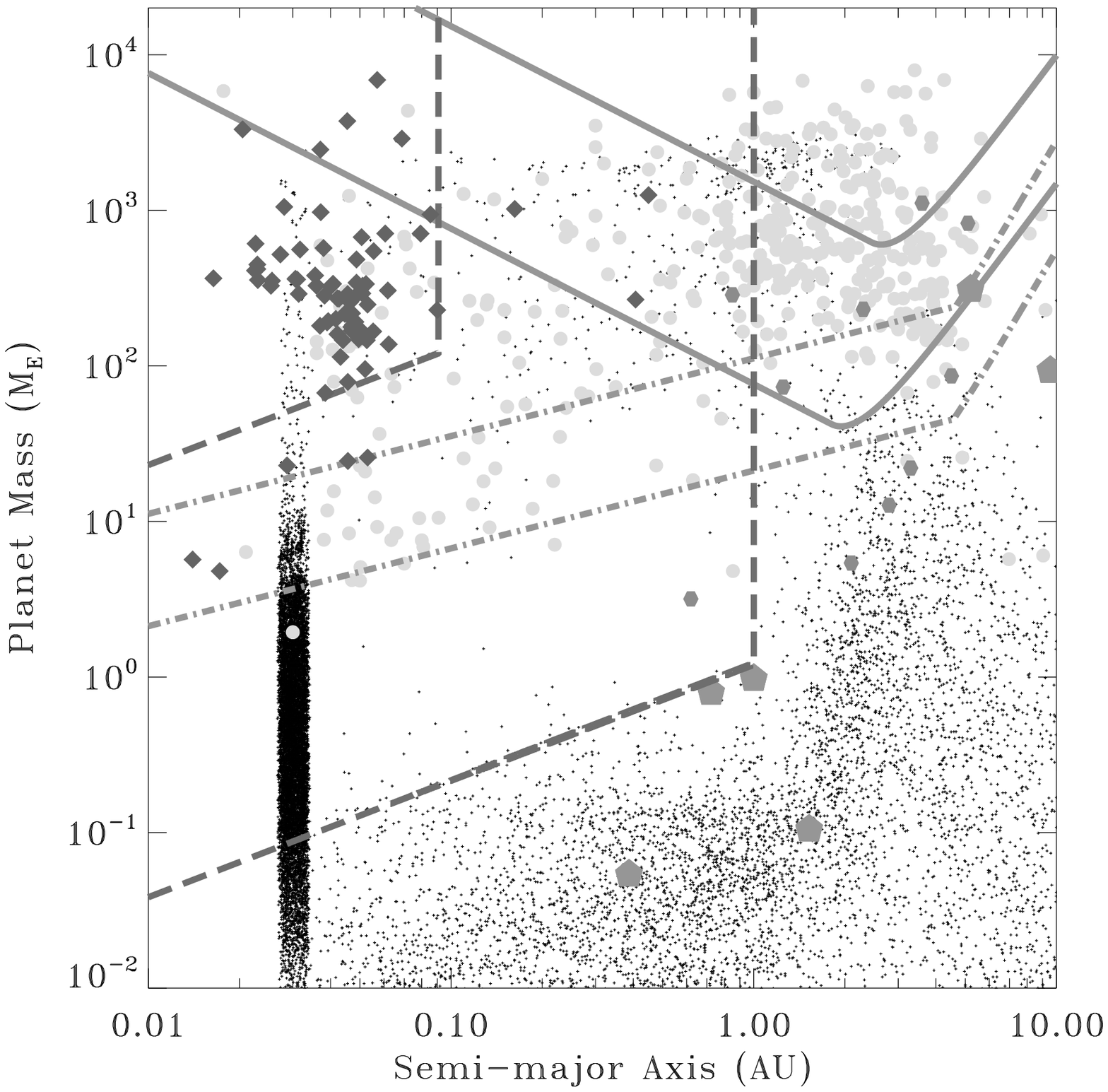} & 
\includegraphics[width=0.47\textwidth]{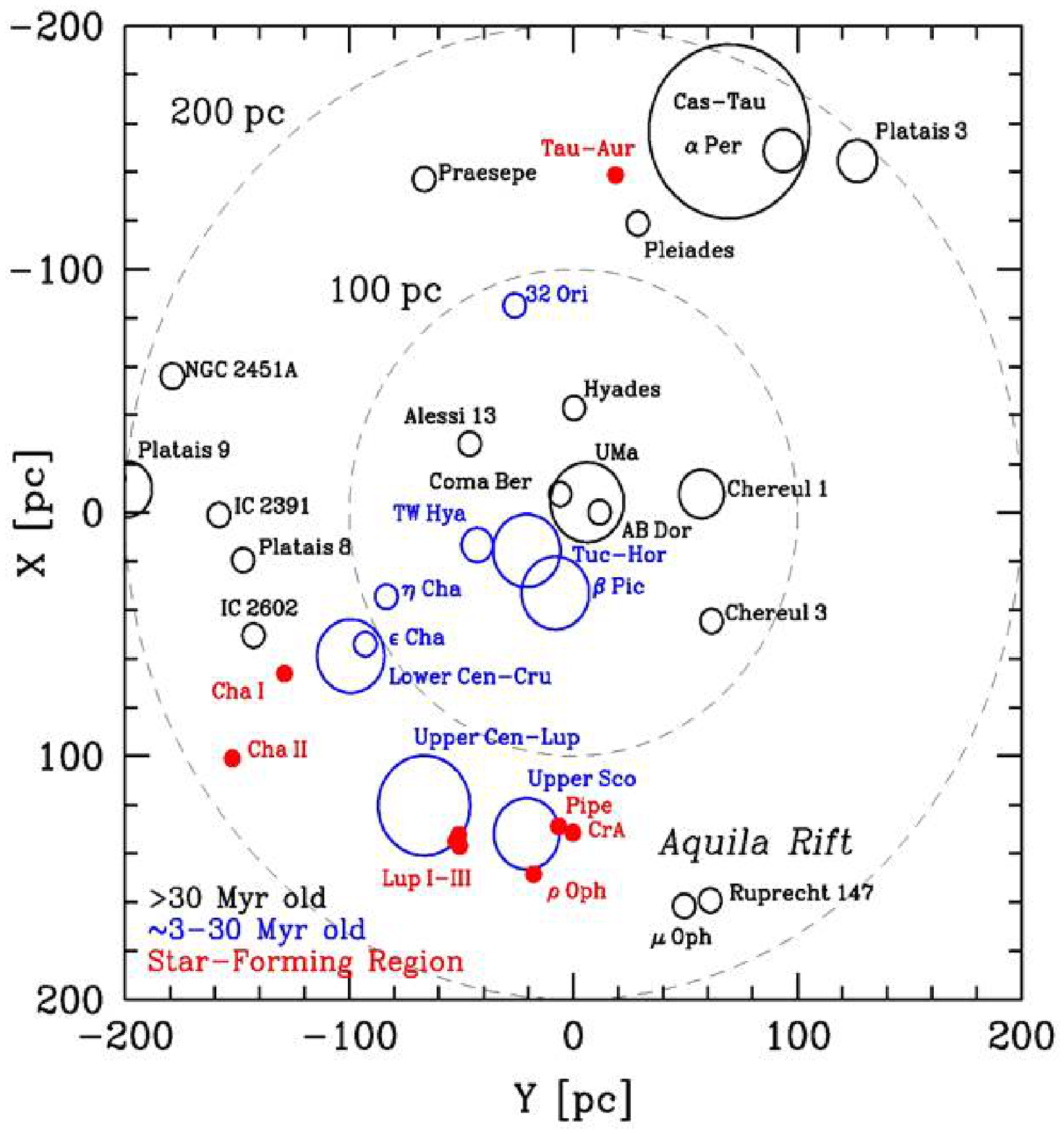} \\
\end{array} $
\caption{{\it Left:} Gaia exoplanets discovery space (solid curves) compared to that of Doppler (dashed-dotted lines) and transit (long-dashed curves) techniques. 
Detectability curves are defined on the basis of a 3-$\sigma$ criterion for signal detection (see Sozzetti 2010 for details). 
The upper and lower solid curves are for Gaia astrometry with $\sigma_\mathrm{A} = 15$ $\mu$as, 
assuming a 1 $M_\odot$ primary at 200 pc and a 0.4 $M_\odot$ M dwarf at 25 pc, respectively, and survey duration set to 5 yr. 
The light-grey filled circles indicate the inventory of Doppler-detected exoplanets as of May 2010. Transiting systems 
are shown as dark-grey filled diamonds. Grey hexagons are planets detected by microlensing. 
Solar System planets are also shown (large grey pentagons). The small black dots represent a theoretical 
distribution of masses and final orbital semi-major axes (Ida \& Lin 2008). 
{\it Right:} The closest ($d< 200$ pc) star forming regions and young stellar kinematic groups (image courtesy of E. Mamajek).
}
\label{fig1}
\end{figure}

The size of the astrometric perturbation $\alpha$, espressed in arcsec, induced on the primary of mass $M_*$ by an orbiting planet of 
$M_p$ (both in $M_{\odot}$) and with a semi-major axis $a_p$ (in AU) scales with the distance $d$ (in pc), to the observer:  $\alpha=(M_p/M_\star)\times(a_p/d)$. 

The sensitivity of Gaia astrometry to (single and multiple) giant planetary companions at intermediate separations around bright, nearby, solar-type dwarfs 
has been the objective of several works in the past (Lattanzi et al. 2000; Sozzetti et al. 2001; Casertano et al. 2008). Those early estimates have been recently 
revisited and complemented by Sozzetti et al. (2014) and Perryman et al. (2014), who used improved (pre-commissioning) knowledge of the 
astrometric error budget and extended the studies to encompass a wider range of primary spectral types and limiting target magnitudes ($G=20$ mag). 
The global figures on which all the above works converge speak of several thousands (possibly $\sim10^4$) astrometrically detectable giant planets 
in the separation range $0.5\leq a\leq 4.5$ AU from their parent stars (see Figure~\ref{fig1}, left panel). The overall all-sky reservoir of stars 
around which Gaia will be sensitive to planetary-mass companions thus exceeds $10^6$. 

As for the population of young stars near the Sun ($d< 200$ pc), there exist of order twenty or so 
nearby star-forming regions, young associations, open clusters and moving groups (see Figure~\ref{fig1}, right panel) with ages in the approximate range $1-100$ Myr (Mamajek, 
this volume; see also Zuckerman \& Song 2004, and references therein, and L\'opez-Santiago et al. 2006, and references therein). 
The likely number of bright ($V< 13-14$ mag) members is on the order of $10^3$ (Mamajek, private communication). 
All these stars will be observed by Gaia with enough astrometric sensitivity to massive giant planets ($M_p\gtrsim2$ $M_J$) orbiting at $2-4$ AU. 
The possibility of detecting giant planets still forming in the protoplanetary disk would constitute fundamental observational evidence to validate the proposed theoretical models of giant planet formation.
It will also probe the trasition region between giant planets and brown dwarfs 
(e.g., Helled et al. 2014, and references therein)
in a regime of orbital separation that would uniquely complement near- and mid-infrared 
imaging surveys (e.g., Burrows 2005, and references therein) for direct detection of young, bright, wide-separation ($a > 30-100$ AU) giant planets, such as those 
presently carried out by SPHERE (Beuzit et al. 2006) and GPI (Macintosh et al. 2014), and in the near future by JWST. 
The above is just one example of the broad range of applications to exoplanets science 
in which Gaia data will act as an ideal complement to (and in synergy with) many ongoing and future observing programs devoted 
to the indirect and direct detection and characterization of planetary systems, both from the ground and in space (Sozzetti 2015). 

\section{On the Synergy Between Gaia Astrometry and Direct Imaging}

\begin{figure}
\centering
$\begin{array}{cc}
\includegraphics[width=0.48\textwidth]{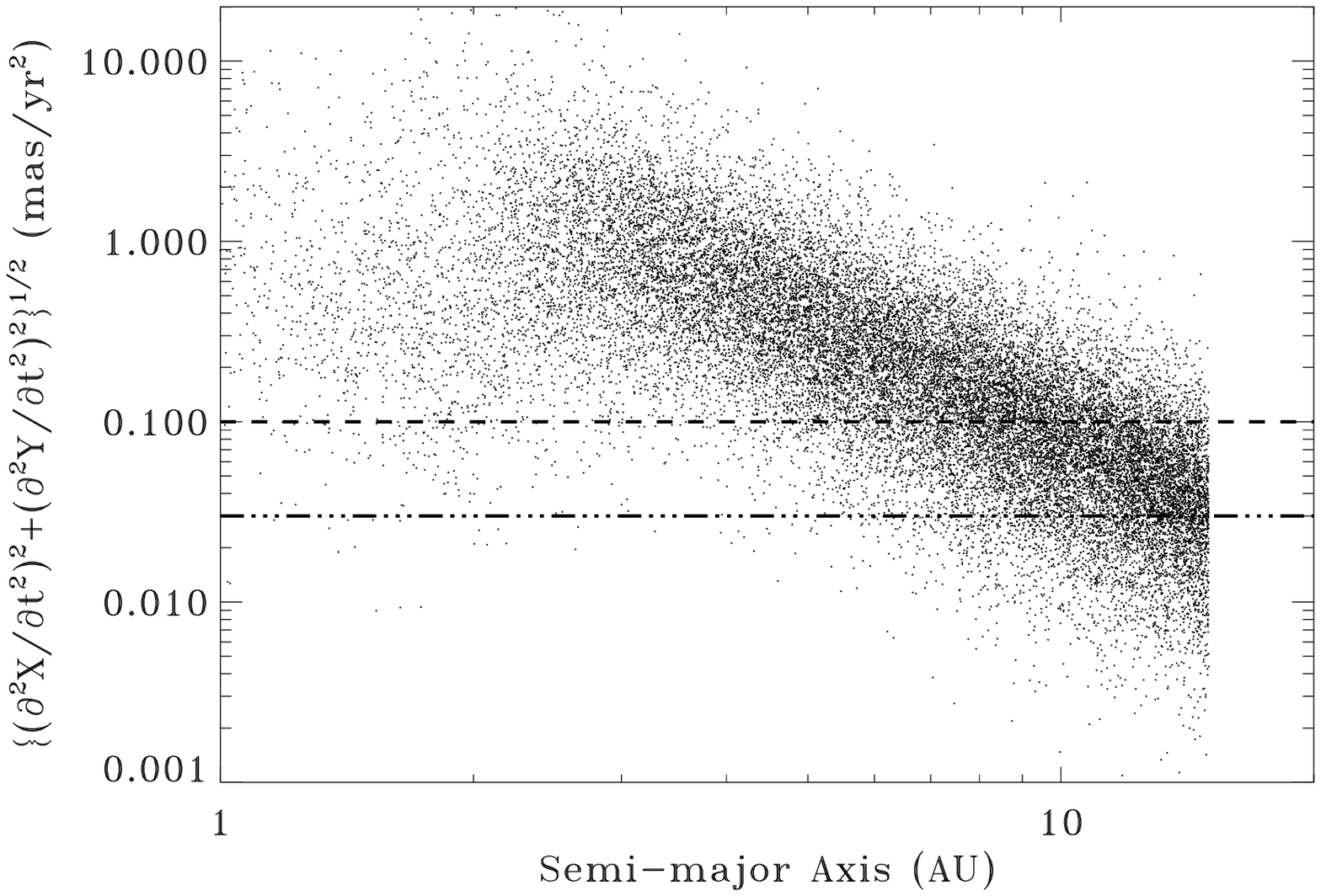} & 
\includegraphics[width=0.50\textwidth]{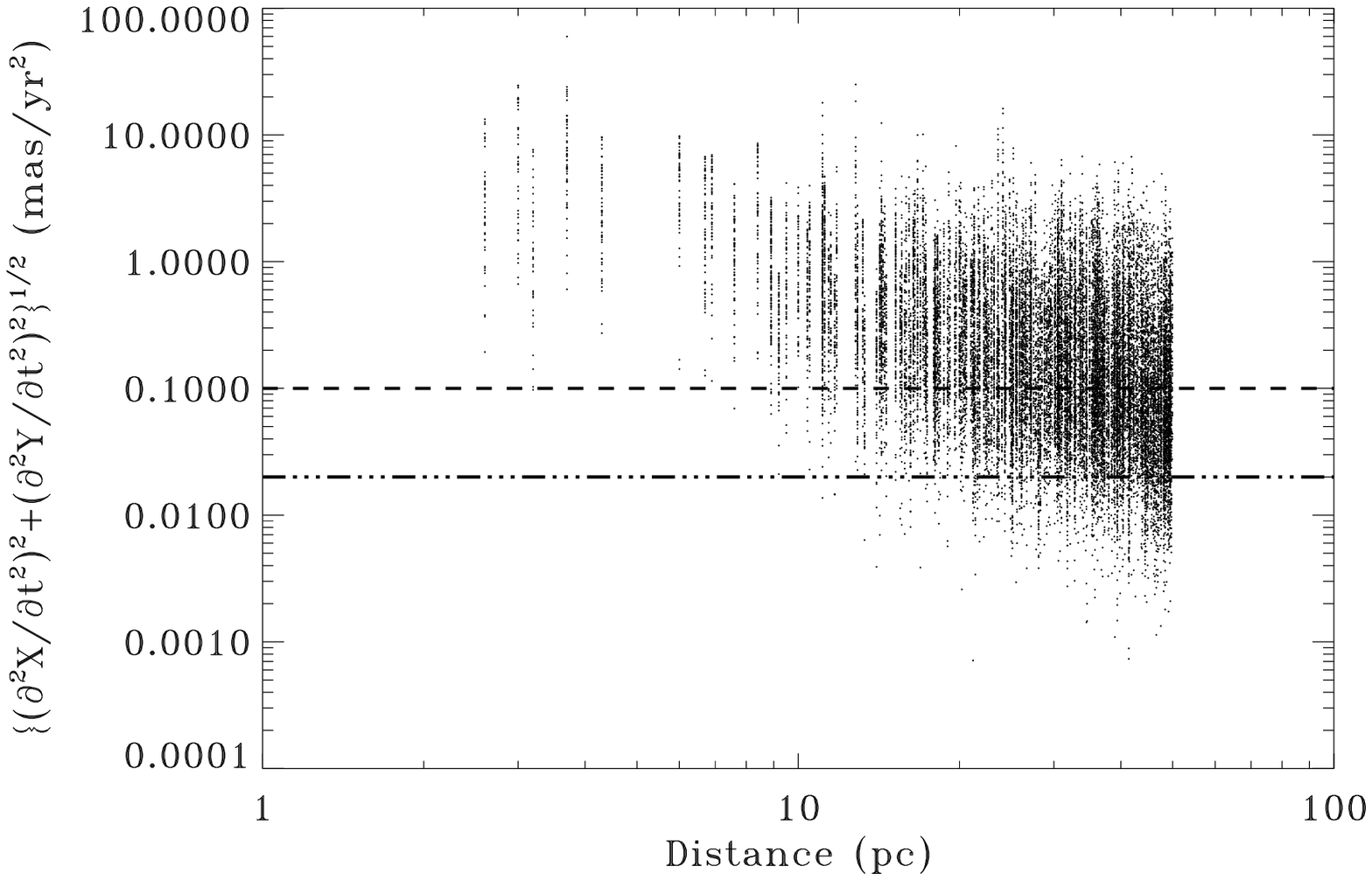} \\
\end{array} $
\caption{{\it Left:} Accelerations in the stellar motion induced by $1-70$ $M_J$ orbiting companions at orbital 
separations $a<15$ AU detected by SPHERE around a sample of $>400$ targets of the GTO program with $V<12$ mag and $d< 50$ pc. 
Dashed and dashed-dotted lines indicate 3-$\sigma_\mathrm{A}$ detection limits with Gaia at mid-mission (2.5 yr) and at mission end (5 yr). Only 5\% of the 
accelerations in Gaia astrometry go undetected at the end of the mission, with a high degree of completeness (99\%) for $a<7$ AU. 
{\it Right:} The same quantity expressed as a function of the distance of the systems from the Sun.
}
\label{fig2}
\end{figure}

The new generation of high-contrast imaging devices (SPHERE, GPI) is primarily sensitive to young giant planets at wide separations. 
Mass estimates of any detected companion heavily rely on structural and evolutionary models, that carry in turn large uncertainties, particularly in the age 
estimates. Dynamical mass constraints are highly desirable, but difficult to obtain. Doppler techniques could help in principle, but a) the amplitudes 
of the radial velocity (RV) signals decrease with orbital period, and b) young stars are trouble (due to rotation and activity). For astrometry, 
the amplitude of the signature increases with $a$, while youth and stellar activity are not an issue for positional measurements with Gaia-like precision. 
It is thus worthwhile to investigate in detail the extents of the synergy between Gaia data and high-contrast imaging programs. We focus here in particular 
on the effectiveness of the combination of SPHERE/VLT direct detections of wide-separation giant planets with Gaia determinations of accelerations in 
the stellar motion due to the orbiting companions.
The aim of this study, which details will be presented in a forthcoming publication (Bonavita et al. 2015, in preparation) is to provide improved constraints on the orbital architecture and mass, thereby helping in the modeling and 
interpretation of giant planets' phase functions and light curves.

\subsection{Simulation Setup} 

The stellar sample used is composed of 439 $V < 12$ mag stars with an age cut-off at $t < 0.5$ Gyr and within 50 pc drawn from the SPHERE GTO sample. 
Synthetic planet populations (1 companion per star) are instead generated with mass and semi-major axis following the distributions from Cumming et al. (2008), 
extrapolated to 70 M$_J$ and 15 AU, and with the other orbital parameters randomly drawn from uniform distributions bracketed by their natural boundaries. 
Then, a large Monte Carlo simulation is run using the MESS code (Bonavita et al. 2012) to identify a statistically significant sample of directly detected systems 
using up-to-date SPHERE detection limits (Zurlo et al. 2014). Next, Gaia-like simulations of the directly imaged systems are generated, using a setup described in Sozzetti et al. (2014). 
Astrometric detection of the companions is based on significant deviations from a five-parameter, single-star model, following which acceleration terms in the stellar motion 
are included in the model to account for curvature effects in the residuals. Finally, the methodology described in e.g. Torres (1999) is adopted to compute the 
cumulative distribution function (CDF) of companion masses inferred by Gaia astrometry that are compatible with the allowed range of orientations and eccentricities, 
and with separation estimates from SPHERE.

\subsection{Preliminary Results}

\begin{figure}
\centering
$\begin{array}{cc}
\includegraphics[width=0.48\textwidth]{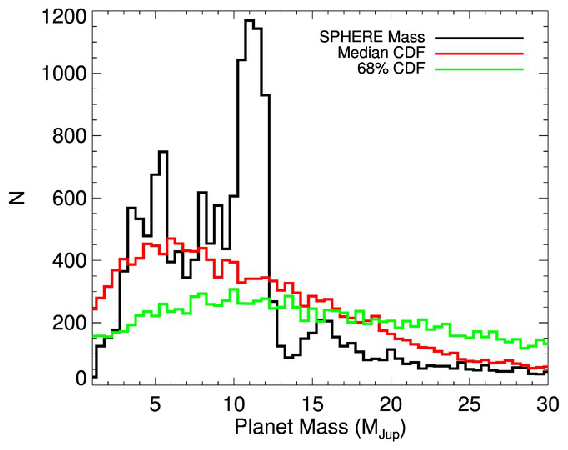} & 
\includegraphics[width=0.48\textwidth]{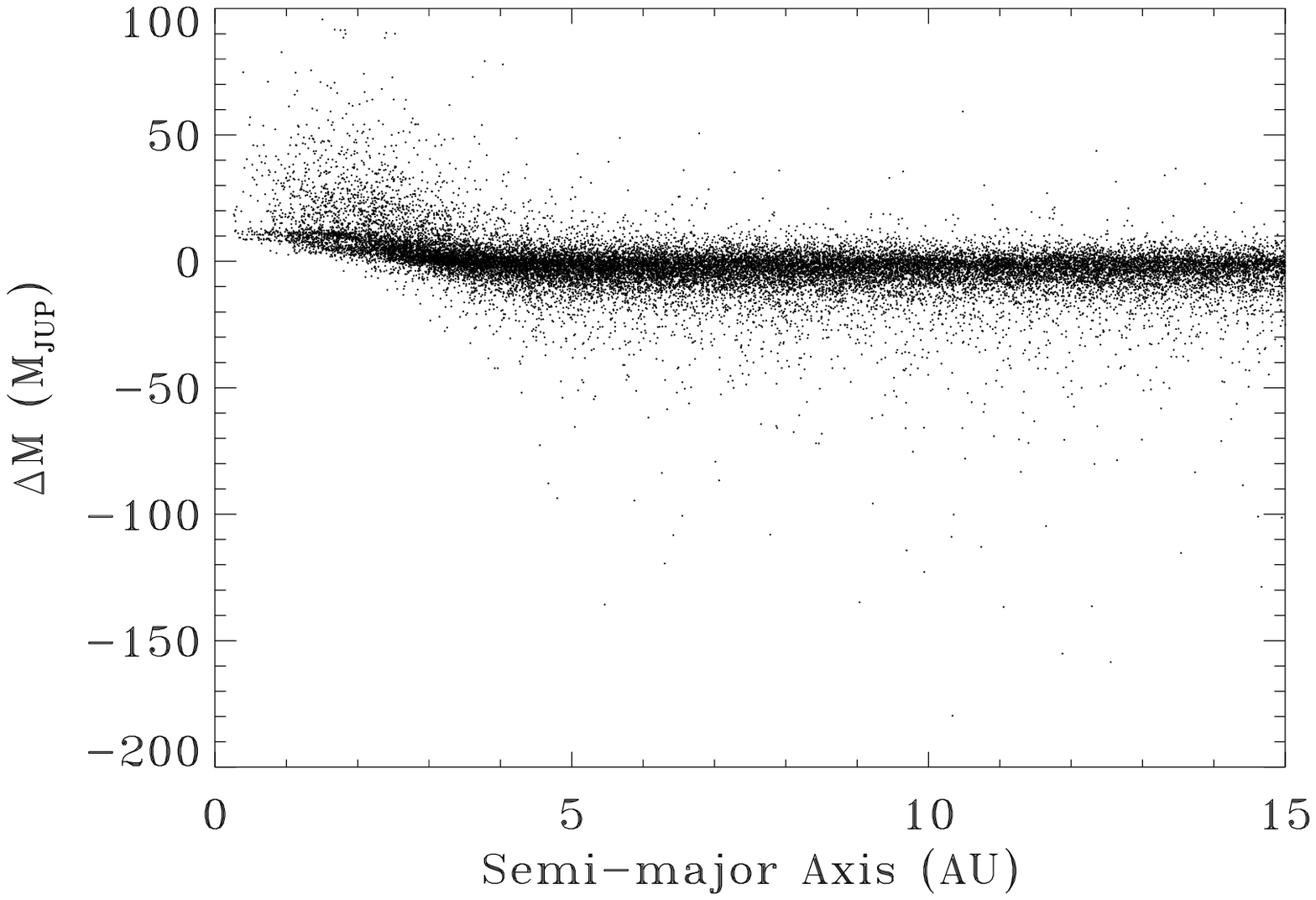} \\
\end{array} $
\caption{{\it Left:} Comparison between the distribution of the nominal companion masses based on SPHERE imaging and the median and 68.3\% credibile 
intervals of the cumulative distribution function of the mass of the companions based on Gaia measurements of accelerations in the astrometry 
of the primaries and the angular separation at a single epoch from SPHERE. 
{\it Right:} Difference between median mass estimate from Gaia astrometry and the SPHERE value as a function of orbital separation.
}
\label{fig3}
\end{figure}

Figure~\ref{fig2} shows the magnitude of the acceleration terms (espressed in mas yr$^{-2}$) estimated by Gaia astrometry for the companions directly imaged 
by SPHERE as a function of the orbital separation (left panel) and system distance from the Sun (right panel). Only $\approx10\%$ of the curvature signals are deemed not significant based on a time baseline 
spanning the nominal 5-yr duration of the Gaia mission. For $a<7$ AU, the completeness levels are higher than 99\%. Similar sensitivity limits apply to systems within $\sim25$ pc 
and to companions with masses larger than $\approx10$ M$_J$ (plot not shown). 

The left panel of Figure~\ref{fig3} shows the comparison between the nominal value of the companion mass as derived from the SPHERE observations (based on inferences from 
theoretical models) and the median and $1-\sigma$ confidence intervals of the mass CDF as obtained from the combination of detections of curvature in Gaia astrometry 
and a measurement of the projected separation from SPHERE (a model-independent estimate). Both the latter metrics tend to overestimate the mass of the companion with respect to the 
one inferred from the models. The systematic effect is likely due to the need (given the range of orbital separations) to fit the Gaia observations with 
improved descriptions of the curvature effects, such as time derivatives of the accelerations. In the right panel of Figure~\ref{fig3} we show instead how the effect reverses when 
orbital periods become comparable to the timespan of the observations: In these cases a full orbital model should instead be fitted to the observations. 

\section{Summary}

The Gaia mission is bound to set the standards in high-precision astrometry for the next decade. Gaia's defining role in the exoplanet arena will be its ability to 
provide a large compilation of new, high-accuracy astrometric orbits of giant planets, unbiased across all spectral types, alongwith exquisitely precise parallaxes. 
There exists a huge synergy potential between Gaia and ongoing and planned exoplanet detection and (atmospheric) characterization programs, 
both from the ground and in space, for much improved understanding of many aspects of the formation, physical and dynamical evolution of planetary systems. We have started 
gauging the effectiveness of the combination of Gaia astrometry and SPHERE high-contrast imaging for constraining masses of directly-imaged companions around 
young stars in the solar neighborhood, as a means to 
reduce inherent degeneracies and uncertainties in model predictions. Our preliminary findings indicate that accelerations in Gaia astrometry for stars with 
companions detected by SPHERE will be easy to spot. We have identified a potential reference metric (the mass CDF) for the assessment of the quality of actual 
mass estimates without relying on model assumptions. Future work will focus on a) quantifying the uncertainties in companion mass estimates when other elements of 
information (e.g., the system age) are factored in, and b) determining the possible improvements in mass determinations when detection of orbital motion is obtained in 
both Gaia and SPHERE data. 

\section*{Acknowledgements}

It is a great pleasure to acnowledge the SOC and LOC of IAU Symposium 314 for organizing a top-class event, spanning a wide range of topics connected to the 
astrophysics of young stars and their planets. This work has been funded in part by ASI under contract to INAF 2014-025-R.0 (Gaia Mission: The Italian Participation to DPAC). 









\end{document}